\documentclass[draft]{agujournal2019}
\usepackage{apacite}
\usepackage{url} 
\usepackage[inline]{trackchanges} 
\usepackage{soul}

\draftfalse

\journalname{JGR: Space Physics}

\begin{document}

%
%

\title{Properties of the Stormtime Plasma Sheet at the Lunar Distance}
%

%
%

\authors{A. Runov \affil{1},
V. Angelopoulos \affil{1},
A. V.  Artemyev \affil{1},
X. An \affil{1}
}

\affiliation{1}{Institute of Geophysics and Planetary Physics, Department of
Earth, Planetary, and Space Sciences,  University of California, Los Angeles,
CA-90095, USA}

\correspondingauthor{Andrei Runov}{arunov@igpp.ucla.edu}

\begin{keypoints}

\item We study two ARTEMIS magnetotail traverses during magnetic storms enabling pre-storm and recovery phase plasma properties comparison.
\item In both cases energization of electrons to relativistic energies and electrostatic noise power enhancement were found.
\item We suggest that local energization via electrostatic turbulence and electron only reconnection may be responsible for the electron energization.

\end{keypoints}

\begin{abstract}
The electron fluxes at energies $E>$100\,keV are shown to be vanishing in the quiet time plasma sheet at geocentric distance of 60 Earth's radii (R$_E$) where the Moon traverses the magnetotail. Fluxes of energetic electrons up to relativistic energies were, however, observed during disturbed space weather conditions.
In this paper, we study the data collected by the two lunar-orbiting  Acceleration, Reconnection, Turbulence and Electrodynamics of Moon's Interaction with the Sun (ARTEMIS) spacecraft during their magnetotail traverses at two magnetic storm events. These observations allow us to compare plasma and field properties obtained at prior to storm and during the storm, including the storm recovery phase. We found that on the storms' recovery phases the average electron temperature increased by a factor of 4 compare to the pre-storm electron temperature. The ion temperature gain, however, did not increase a factor of 2. That leads to a decrease of ion to electron temperature ration to $\langle{T_i}/{T_e}\rangle\approx$3, in contrast to the pre-storm value of 7 to 9. We also found an increase in integral power of electrostatic fluctuations up to $\approx$2\,|mV/m|. Our observations suggest that the electrons were energized to energies $E>$100\,keV in the magnetotail. Although the exact mechanism of this energization remains unclear, we suggest that energization via continuous sporadic electron-only reconnection associated with electrostatic turbulence may be responsible for the anomalous electron energization.
\end{abstract}

%
%

%


%
%

\section{Introduction}
The Moon traverses the nightside of the Earth's magnetosphere - the magnetotail - at geocentric distances $R\sim$60 Earth's radii ($R_E$) during 3 to 4 days every month. The near-Moon magnetotail serves as a unique plasma physics laboratory to study plasma behavior in a weak and highly fluctuating magnetic field. In many respects, the  magnetic conditions in the near-Moon magnetotail are opposite to those in the inner magnetosphere: the magnetic field strength during quiet conditions does no exceed 10\,nT; the equatorial (i.e. at $|B_x|\approx$0) magnetic field is about 1\,nT without a preferential direction ($B_y$ and $B_z$ are equiprobable), and magnetic fluctuation level $\delta{B}/\langle{B}\rangle\rightarrow$1 \cite<e.g.,>[and references therein]{runov23}. At the same time, a current sheet with the current density up to 10s\,nA/m$^2$ and thickness of 10$^3$\,km, which is comparable to an ion thermal gyroradius or a few ion inertial lengths, is embedded into the plasma sheet in the lunar-distant magnetotail \cite<e.g.,>{kamaletdinov24, kamaletdinov25GRL}. The magnetic field and current sheet geometry of the  lunar-distant magnetotail depend on solar wind and interplanetary magnetic field (IMF) conditions. The Earth's dipole does not play a significant role.

Similar to the magnetic field, plasma conditions in the near-Moon magnetotail are quite different from those in the inner magnetosphere and in the near-Earth plasma sheet. The average energy of ions and electrons in the quiet-time lunar-distant plasma sheet depends on the solar wind kinetic energy \cite{runov23}. The average ion and electron temperatures are of 0.7\,keV and 0.1\,keV, respectively. During low geomagnetic activity,
ion and electron fluxes at energies exceeding $\sim$20\,keV are negligible. The ion and electron distributions are, even during the low activity and absence of fast plasma flows, are non-Maxwellian, showing a pronounced high-energy tail \cite{runov23}.
The ion and electron fluxes at 20\,keV range  may increase by a factor 5 to 10, compared to quiet-time levels, within tailward rapid flux transport (RFT) events with the flux transport rate $E_y=({\bf v}\times{\bf B})_y>$2/,mV/m \cite{runov18}. However, $\geq$100\,keV fluxes in the RFTs, on average, barely exceed the instrument noise level.

On the other hand, fluxes of energetic, even relativistic (up to 200\,keV) electrons have been observed in the mid and distant magnetotail by IMP~7, ISSI-3, and Geotail \cite<e.g.,>{sarafopoulos04, sarris76, sarris96, baker77, richardson96}. The observed energetic electron flux events fell into two categories: field-aligned beams and isotropic omnidirectional fluxes, with the latter observed primarily on the closed magnetic field lines \cite{baker77}.

Recently, the Moon-orbiting Acceleration, Reconnection, Turbulence, and Electrodynamics of Moon's Interaction with the Sun (ARTEMIS, \citeA{angelopoulos11}) spacecraft detected prolonged (up to 50 hours) enhancement in relativistic ($E>$100\,keV) electron fluxes in the postmidnight magnetotail during magnetic storm recovery phases \cite{runov25b}. The relativistic electron fluxes were observed by both ARTEMIS probes situated earthward as well as tailward of the Moon. Simultaneously, relativistic electron flux enhancements were detected by the Magnetospheric Multiscale (MMS) spacecraft in the premidnight magnetotail at $R \approx$20\,R$_E$ (see Fig.~1 in \citeA{runov25b}). Simultaneous observations at L1 by the Wind spacecraft did not reveal a significant increase in a level of energetic electron fluxes in the solar wind. These observations suggest that the electrons were energized up to relativistic energies internally in the magnetosphere.

What may be a source of the relativistic electrons observed at lunar orbit? What are mechanisms of their energization up to relativistic energies?

Energetic (up to $\sim$100\,keV) electrons were frequently observed in the magnetosheath \cite<e.g.>[and references therein]{cohen17}. Some studies attributed them to acceleration of suprathermal solar wind population by whistler waves \cite<e.g.>{oka06, wilson12GRL}, or via the drift shock acceleration \cite<e.g.,>{liu22ApJ}. Others argue that these electrons are of magnetospheric origin and leak from outer radiation belts \cite<e.g.>{cohen17}. Regardless of the origin, there were no studies discussing a possibility of energetic magnetosheath electrons to entry to the mid-distant magnetotail.

The radiation belts are the main reservoir of relativistic electrons in the magnetosphere. Numerous studies showed that hundreds keV electrons often leak from the outer radiation belt through the magnetopause to magnetosheath \cite<e.g.,>[and references therein]{cohen17}. Modern comprehensive test-particle simulations showed that the leaking outer radiation belt particles may be involved in a complex motion along the magnetopuase and reenter the nightside magnetosphere \cite{sorathia17, cohen21}. However, the question whether these particles may reenter magnetosphere in mid-distant tail to be detected at lunar orbit remains unanswered.

Some studies also suggest that outer radiation belt particles may scape through the cusp and populate the plasma sheet boundary layer \cite<e.g.,>{delcourt99}. Although the analysis showed a possibility for these particles to enter the plasma sheet, it is questionable that they may reach lunar orbit.

In the mid and near-Earth magnetotail, enhancements in energetic ($E\geq$100 keV) electron fluxes are associated with rapid flux transport events (RFTs, defined by $E_y>2$\,mV/m, see \citeA{schodel01a}) and localized dipolarizations \cite<e.g.,>{birn14, runov15, gabrielse14, gabrielse16, sorathia18}. Statistically, electron fluxes above the instrument noise level have been detected at energies up to 300\,keV \cite{gabrielse22SW}. Observations have also reported relativistic electron fluxes during magnetic storms and high-intensity, long-duration continuous $AE$ activity (HILDCAA) events \cite<>{angelopoulos19, runov25a}. Strong enhancements in $\geq$100 keV electron fluxes have been observed during in situ measurements of magnetotail reconnection \cite<e.g.,>{fujimoto01, oieroset02, turner21b, oka22POP}.
Observations by the Magnetospheric Multiscale (MMS) mission at geocentric distances $R>$\,20\,R$_E$ indicate that, at times, sufficient flux exists in the central plasma sheet to provide a source of $>$1\,MeV electrons for the outer radiation belt \cite{turner21a}. Furthermore, recent remote sensing studies using low-Earth orbiting satellites have detected bursts of electron fluxes up to $\approx$3\,MeV poleward of the isotropy boundary (IB), which marks the inner edge of the plasma sheet \cite{xjzhang25}. However, whether these energetic electrons reach lunar orbit remains an open question.

In this paper we extend our analysis of ARTEMIS observations during the two magnetic storms on August 2018 and on September 2024 during which ARTEMIS probes traversed magnetotail from west to east \cite{runov25b}. These observations allow us to follow the storm-time evolution in magnetic field and plasma parameters and compare properties of pre-storm and recovery phase plasma sheet. These observation may shed new light on the nature of  enigmatic relativistic electron populations detected in lunar-distant magnetotail.

\section{Data Analysis}
\subsection{The Datasets}
We use data from the two ARTEMIS probes during their
traversals of the magnetotail at $R\approx$60\,R$_E$.
The probes are in 26\,hr period elliptical equatorial orbits with $\sim$100\,km x 19,000\,km altitude.

We use 0.25\,s-resolution magnetic field measurements from  the THEMIS Fluxgate Magnetometer \cite<FGM,>{auster08}
and particle measurements from the THEMIS Electrostatic Analyzer
 \cite<ESA,>{mcfadden08} and the THEMIS Solid State Telescope \cite<SST,>{angelopoulos08a} at energies of $\sim$10\,eV to 25\,keV and 30\,keV to 720\,keV, respectively. We use combined ESA and SST distributions \cite{runov15} to calculate the bulk velocity and the electron omnidirectional differential fluxes and pitch angle distributions in the 100\,eV to 720\,keV energy range. We also use measurements from the THEMIS Electric Field Instrument \cite<EFI>{bonnell08} and the Search Coil Magnetometer \cite<SCM>{roux08}.

We use the OMNI database for the solar wind and IMF parameters and for geomagnetic indices.

\subsection{Events Overview}
\begin{figure}[t]
\includegraphics[width=\textwidth]{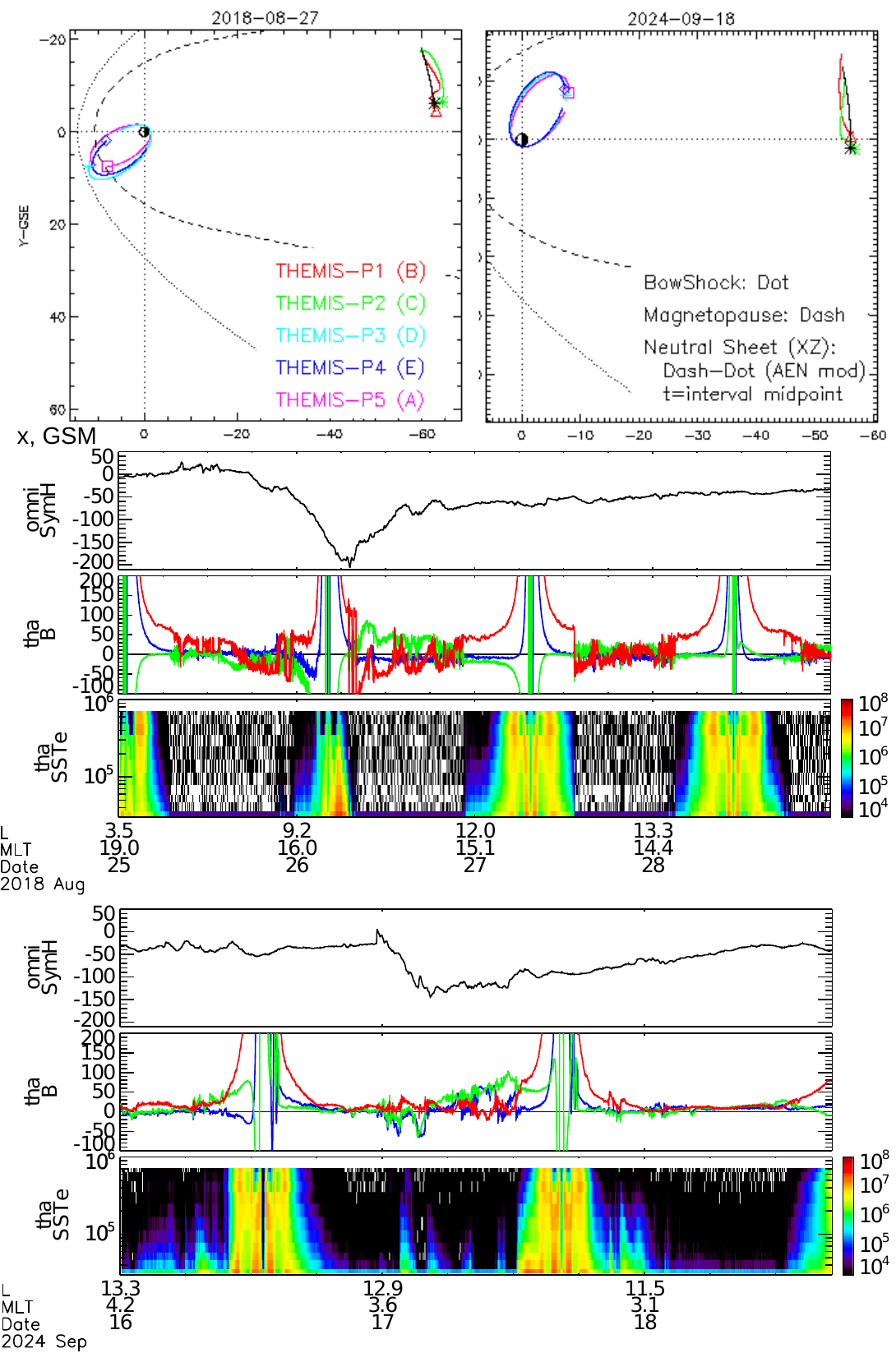}
\caption{\label{thasplot}
THEMIS and ARTEMIS probe trajectories, OMNI ${SymH}$ index, and THEMIS A FGM and SSTe data for Event 1 and 2.
}
\end{figure}
Prior to discuss ARTEMIS observations in the lunar-distant magnetotail, we show THEMIS and ARTEMIS probe trajectories and brief summaries of observations by Earth orbiting THEMIS probe A during both storm events in Figure~\ref{thasplot}. During Event~1 THEMIS apogee was in the post-noon sector. The THA FGM data indicates that the probe sampled the magnetosheath in between diving into the inner magnetosphere. No electron fluxes above the instrument noise level were detected by the THA SSTe during the magnetosheath excursions. Conversely, in Event ~2 THEMIS apogee was in the dawn sector. Strong electron fluxes in the SST ($E>30$\,keV) range were detected in the dawn plasma sheet prior to the storm, during the storm's main phase, and during the recovery phase.

\begin{figure}[t]
\includegraphics[width=\textwidth]{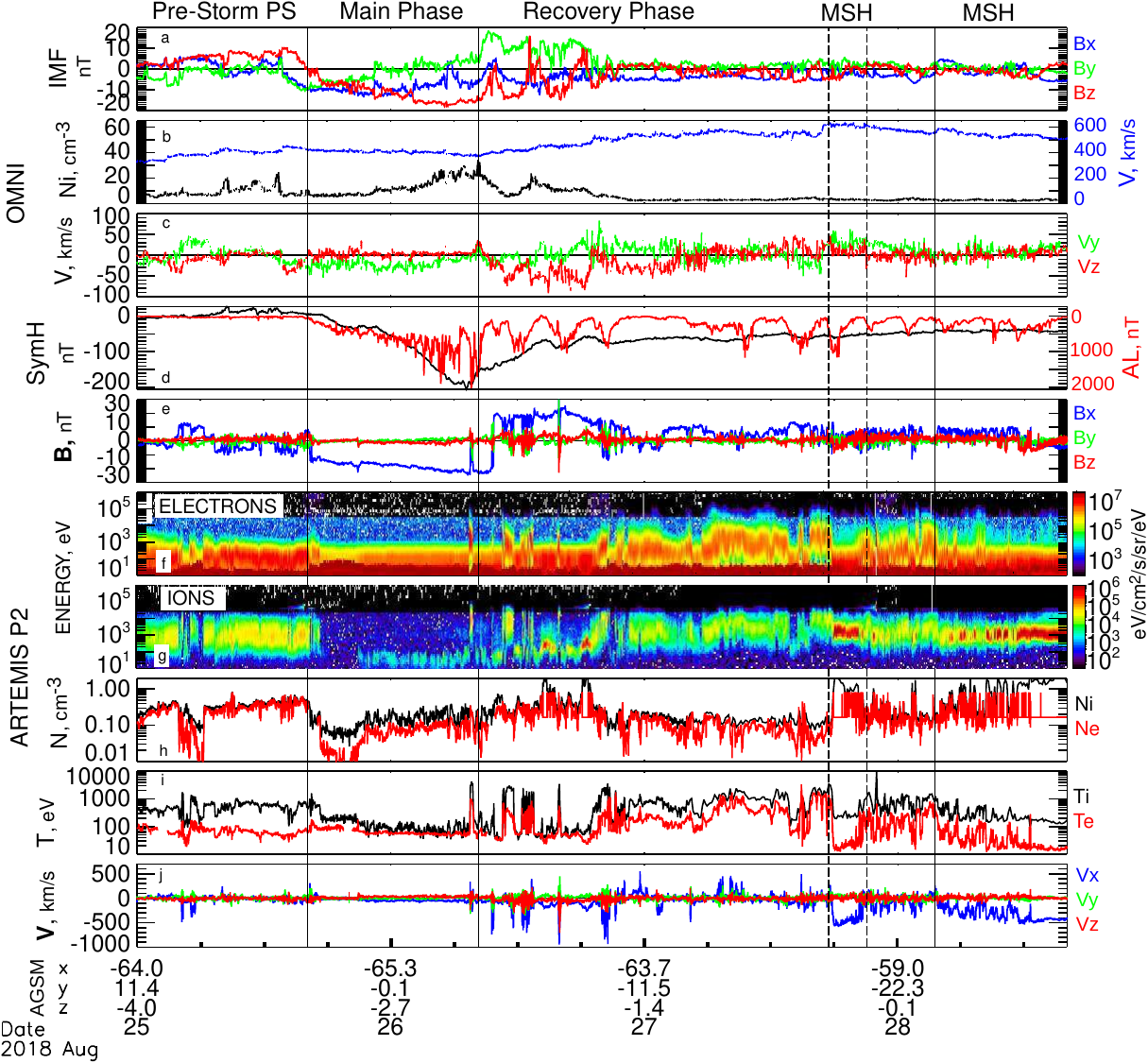}
\caption{\label{artm_omni_18_overview}
The August 2018 storm event overview:
a) IMF components, b) solar wind density (black) and speed (blue), c) $y$ (green) and $z$ (red) solar wind velocity components, d) $SymH$ (black) and $AL$ (red) indices from OMNI; e) magnetic field components, f) electron and g) ion energy time spectrograms, h) electron (red) and ion (black) densities, i) electron (red) and ion (black) temperatures, and bulk velocity components from ARTEMIS P1 during the magnetotail traversal between 25 and 28 August 2018.
}
\end{figure}

Figure \ref{artm_omni_18_overview} shows summary of solar wind/IMF and magnetotail magnetic field and plasma parameters changes during the August 2018 storm event. The storm main phase started at about 1740\,UT on 25 August (first vertical line in Fig.~\ref{artm_omni_18_overview}) after IMF turned southward around 1600\,UT. The main phase was preceded by an increase in $SymH$ caused by increases in solar wind speed and density. At that time ARTEMIS P2 was in the dusk magnetotail sector, travelling between $Y_{AGSM}\approx$11\,R$_E$ and $Y_{AGSM}\approx$4\,R$_E$. The ARTEMIS observations during the pre-storm interval show typical quiet plasma sheet (PS) parameters with the magnetic field $x$-component fluctuating between $\pm$10\,nT and ion and electron temperatures of 1\,keV and 0.1\,keV, respectively \cite{runov23}.

\begin{figure}[t]
\includegraphics[width=\textwidth]{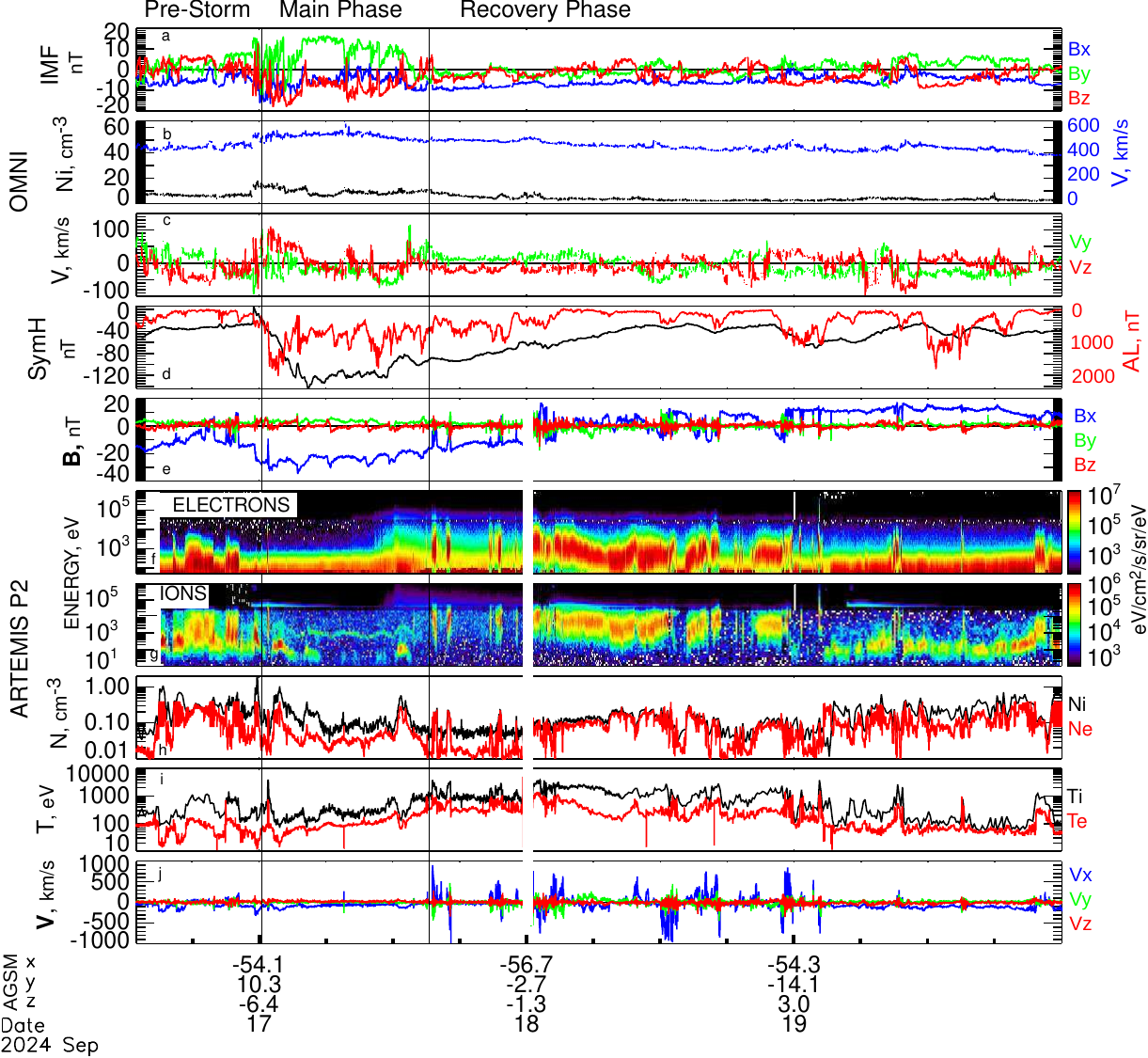}
\caption{\label{artm_omni_24_overview}
The September 2024 storm event overview. The same format as in Fig.~\ref{artm_omni_18_overview}
}
\end{figure}

In both cases ARTEMIS probes were at high $|B_x|$ level, i.e., far from the equatorial plane during the storms' main phase. Only low-energy ion populations (plasma mantle and/or ionospheric outflow) and low-energy electrons on top of spacecraft potential level (saturated stripe in electron energy-time spectrograms) were observed. Although the low energy ion population is interesting, it is beyond the scope of this paper. In further analysis we focus on comparison between observations during pre-storm interval and on the st0rms' recovery phase.

\subsection{Recovery Phase PS Properties}
The ARTEMIS observations during these two storm events provide an opportunity to compare pre-storm and storm recovery phase plasma sheet properties. A caution is needed, however. The ARTEMIS probes were at the dusk flank of the magnetotail during the pre-storm interval and traversed the postmidnight sector towards the dawn magnetitail flank during recovery phase.

Below we show statistical distributions of fields and plasma parameters observed during pre-strom intervals and during recovery phases of August 2018 (Event~1) and September 2024 (Event~2) magnetic storms. The exact UTs of pre-storm and recovery phase intervals for Events~1 and 2 were

\noindent
\small{
\begin{tabular}{lll}
 Event & Pre-storm & Recovery \\
 E1 & 2018-08-25/06:00:00 - 2018-08-25/16:00:00 & 2018-08-26/20:37:00 - 2018-08-27/17:40:00 \\
 E2 & 2024-09-16/17:30:00 - 2024-09-16/20:00:00 & 2024-09-18/00:30:00 - 2024-09-18/18:00:00. \\
\end{tabular}
}

\begin{figure}[t]
\includegraphics[width=\textwidth]{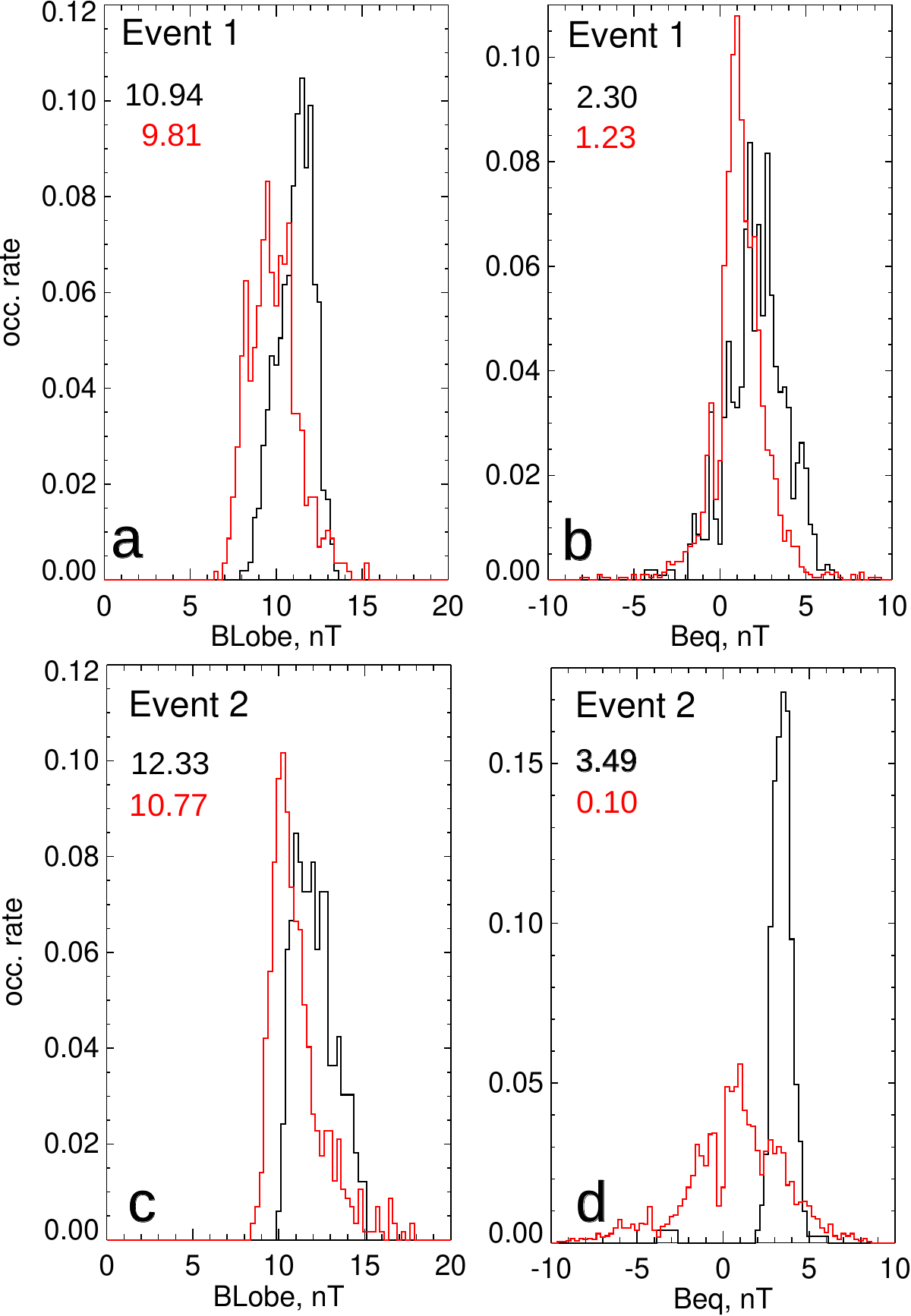}
\caption{\label{bl_beq}
Statistical distributions of lobe magnetic field strength ($B_L$, a, c)  and equatorial magnetic field
($B_{eq}$, b, d) observed during the pre-storm interval (black) and during the recovery phase (red) in Event~1 (upper panel) and Event~2 (bottom panel). The corresponding median values (in nT) are printed in corresponding colors.
}
\end{figure}

Figure \ref{bl_beq} shows distributions of the lobe magnetic field strength $B_L$ (panels a and c) and the equatorial magnetic field $B_{eq}=\sqrt{B_y^2+Bz^2}sign(B_z)$ within $|B_x|<$1.0\,nT (panels b and d) for pre-storm intervals (black curves) and recovery phases (red curves) for Event~1 and Event~2, respectively.
Evidently, the lobe field strength decreases during the recovery phase with respect to the pre-storm values by a factor of 0.1. The equatorial magnetic field was dominantly positive during pre-storm intervals for both events. In Event~1, the recovery phase $B_{eq}$ evidently shifted to smaller values saying dominantly positive. In Event~2 recovery phase, $B_{eq}$ was broadly distributed, roughly equally in positive and negative values, with the median close to zero.

\begin{figure}[t]
\includegraphics[width=\textwidth]{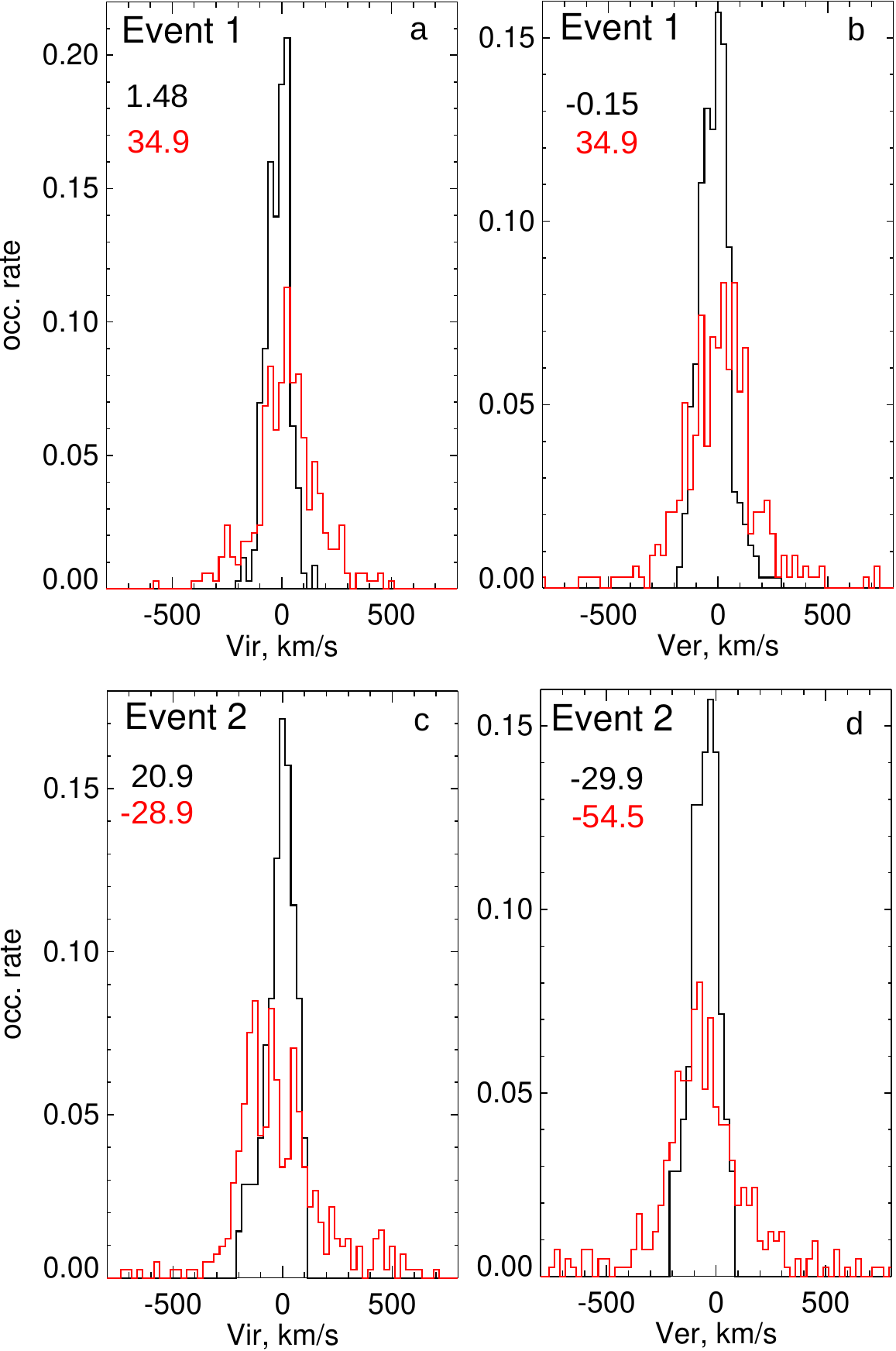}
\caption{\label{artm_vie}
Ion (a, c) and electron (b, d) radial velocities distributions during the pre-storm interval (black) and the recovery phase (red) in Event~1 (a, b) and Event~2 (c, d). The corresponding median values (in km/s) are printed in corresponding colors.
}
\end{figure}

Figure \ref{artm_vie} shows distributions of ion and electron radial bulk velocities
$V_r$=$\sqrt{V_x^2+V_y^2}$
calculated for Event~1 and 2 pre-storm and recovery phase intervals. In Event~1, both ion and electron pre-storm velocities were close to zero with a narrow statistical distributions. Distributions for the recovery phase are significantly broader, with significant counts up to $\pm$500\,km/s, although the median values for both $V_i$ and $V_e$ were as low as $\approx$35\,km/s. In Event~2 pre-storm velocities distributions were similar to those in Event~1, although the median values were by a factor of 10 larger. The recovery phase velocities in Event~2 were broader distributed, compare to those in Event~1 with the median shifted to negative values. In general, distribution of ion and electron velocities in both events were very similar with a longer, mainly positive tail on the recovery phase.

\begin{figure}[t]
\includegraphics[width=\textwidth]{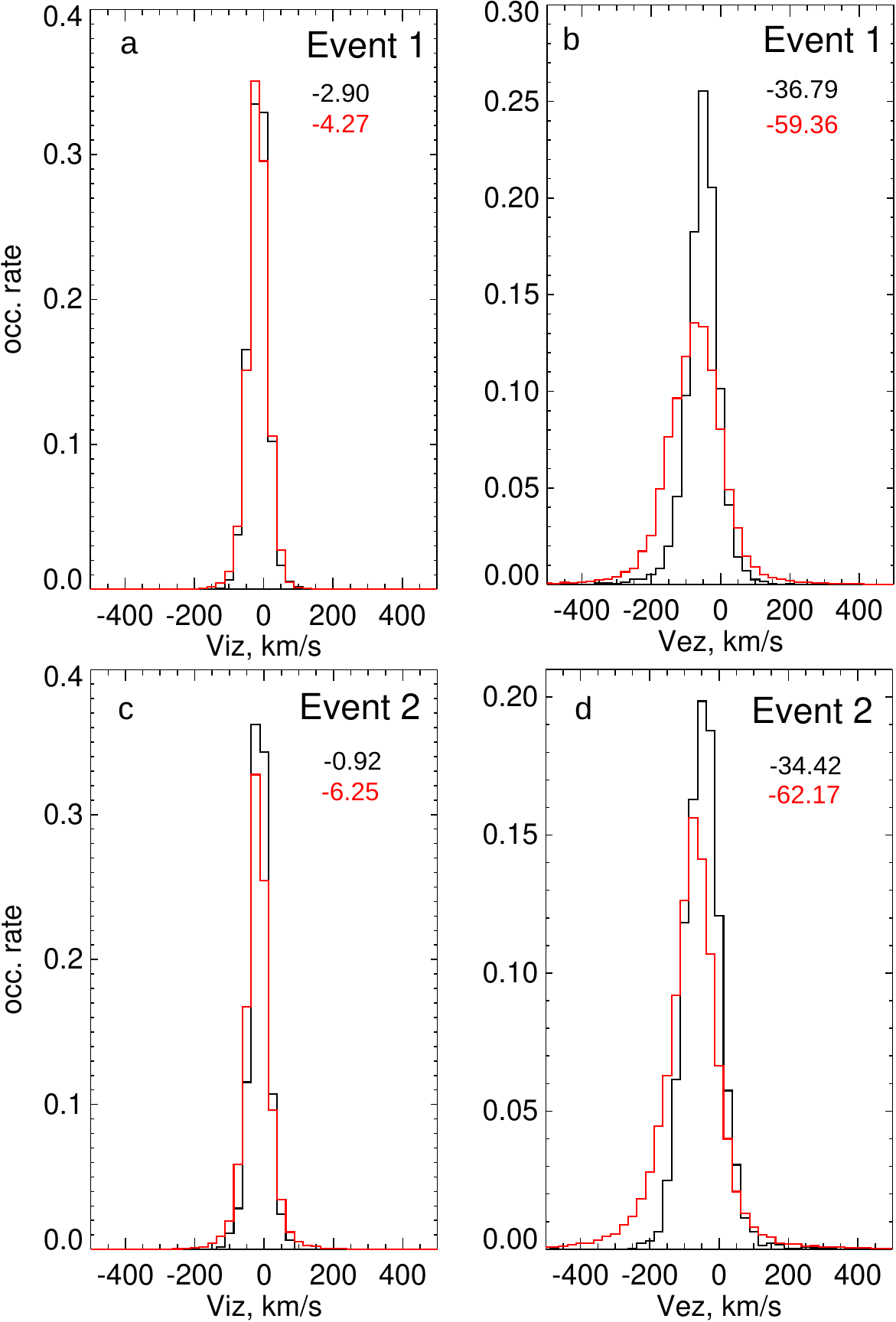}
\caption{\label{artm_viez}
Ion (a, c) and electron (b, d) polar velocities distributions during the pre-storm interval (black) and the recovery phase (red) in Event~1 (a, b) and Event~2 (c, d). The corresponding median values (in km/s) are printed in corresponding colors.
}
\end{figure}

Figure \ref{artm_viez} shows distributions of the polar velocity (i.e., along $z_{GSM}$) of ions and electrons during pre-storm and recovery intervals. Evidently, in both events changes in the ion polar velocity were negligible, whereas electron polar velocity distributions at the recovery phase exhibit skewness towards negative values. Thus, because in both events probes were, predominantly, in the northern hemisphere ($B_x>0$), the negative electron velocity means an inflow towards the magnetic equator.


\begin{figure}[t]
\includegraphics[width=\textwidth]{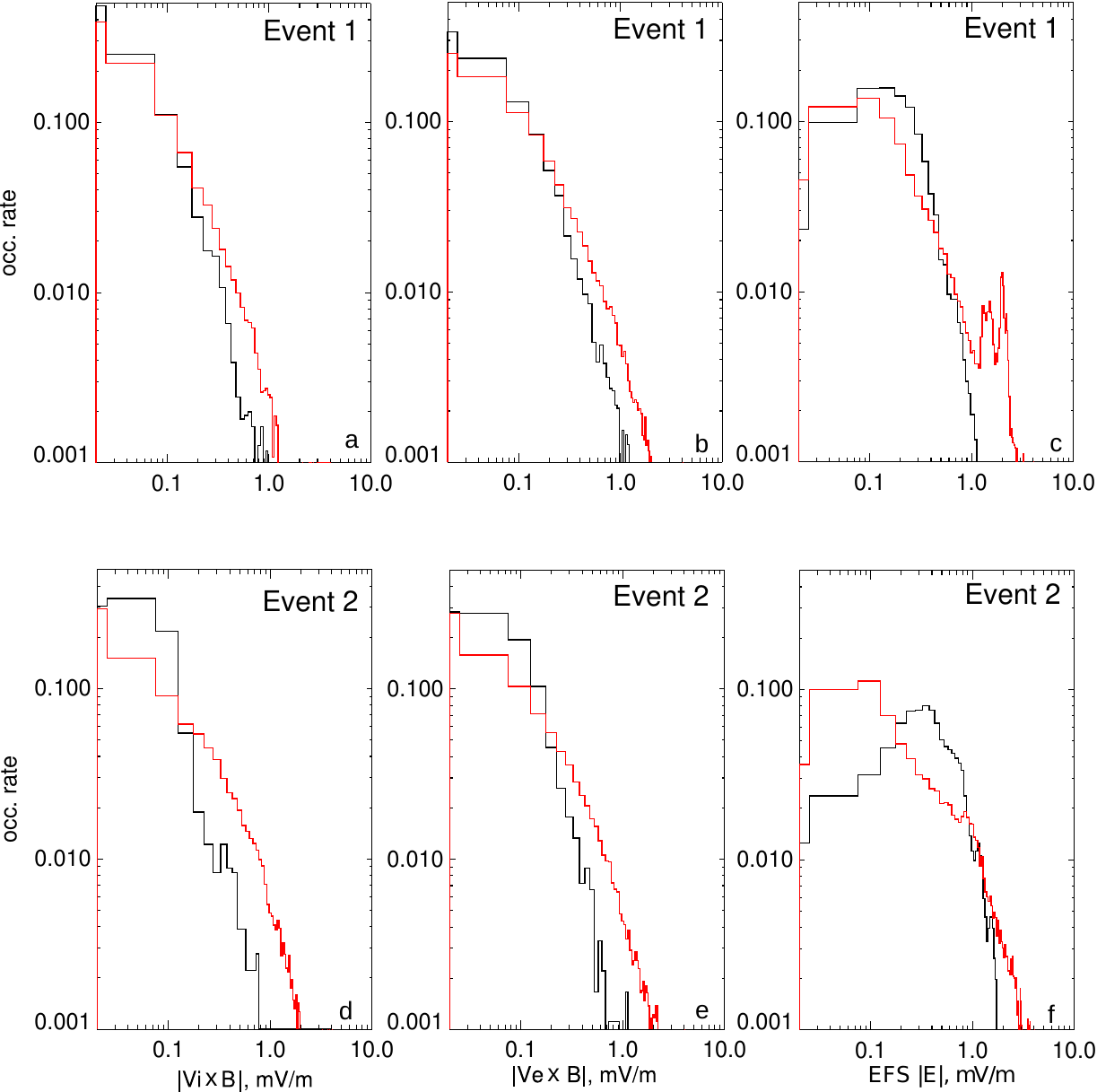}
\caption{\label{artm_E}
Distributions of absolute values of ${\bf v}_i\times{\bf B}$ (a, d),
${\bf v}_e\times{\bf B}$ (b, e), and the electric field ($E$) measured by the EFI instrument during the pre-storm interval (black) and the recovery phase (red) in Event~1 (a, b, c) and Event~2 (d, e, f).
}
\end{figure}

Figure \ref{artm_E} shows distributions of the magnetic field transport rates, calculated using ion (panels a and d) and electron (panels b and e) bulk velocities, respectively, as
$E=\sqrt{
({{\bf V}_j}\times{\bf B})_x^2 +
({{\bf V}_j}\times{\bf B})_y^2}$, where $j=i,e$, for pre-storm and recovery phase intervals. Panels c and f show distributions of the DC electric field $E=\sqrt{E_x^2+E_y^2}$ measured by the EFI instrument during pre-storm and recovery phase intervals. In Event~1, the changes in the magnetic flux transport by ion velocity were minor, the $|{bf V}_i \times {\bf B}|\leq$1\,mV/m. The $|{bf V}_e \times {\bf B}|$ distribution show somewhat broader distribution, with a significant occurrence rate at 1$\leq|{bf V}_e \times {\bf B}|\leq$2\,mV/m.
The measured electric field distribution exhibit the second peak around 2\,mV/m during the recovery phase,  which, likely, indicate that kinetic effects were in action. In Event~2, in contrast, the magnetic field transport rates, calculated with $V_i$ and $V_e$, and the measured electric field are in agreement and reached 2\,mV/m.

\begin{figure}[t]
\includegraphics[width=\textwidth]{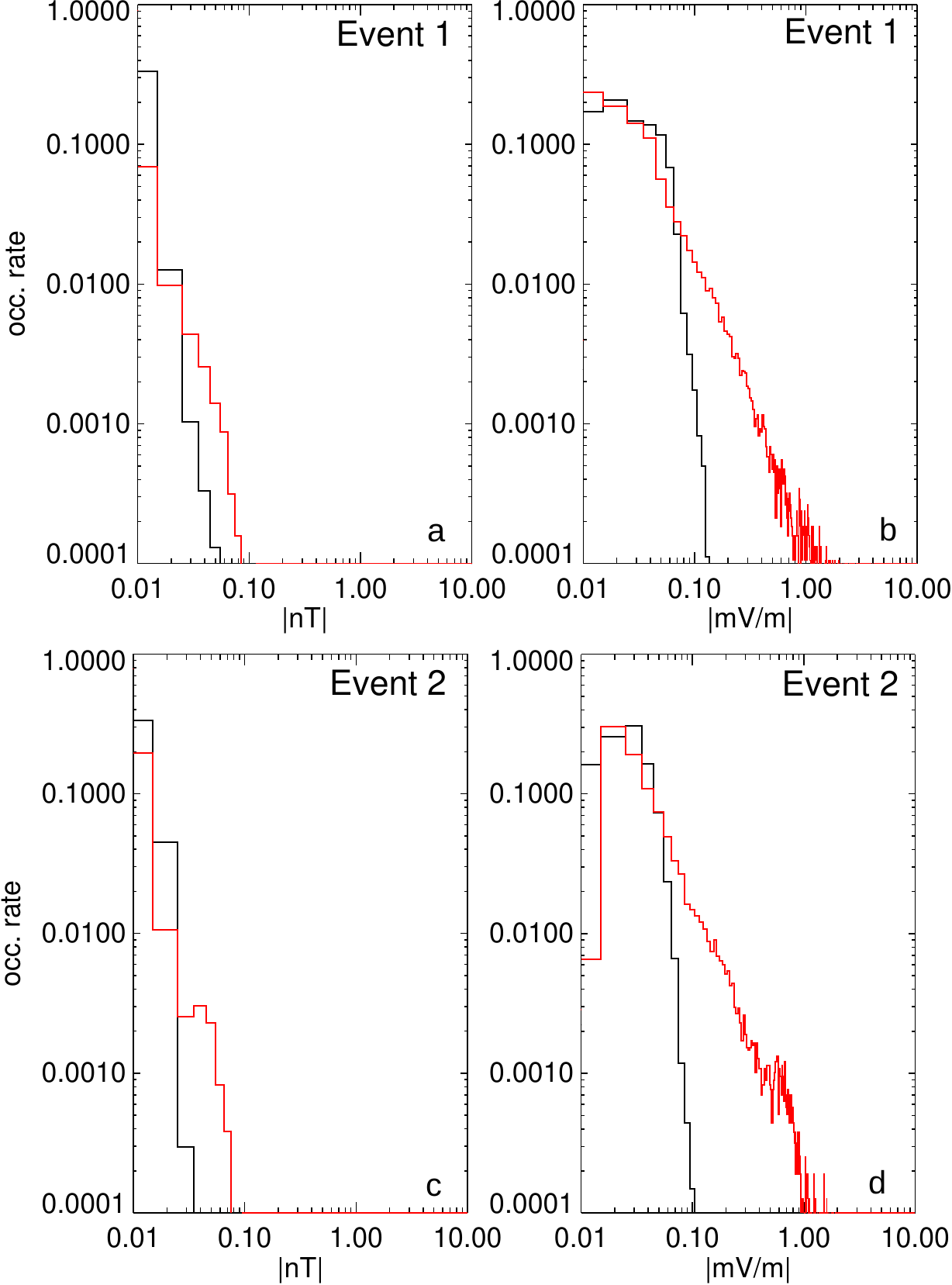}
\caption{\label{artm_FBK}
Distributions of the FBK wave power integrated over frequency bands [9.05, 36.20, 144.2, 572.0 2689.0]\,Hz from SCM (in nT) and from EFI (in mV/m)  during the pre-storm interval (black) and the recovery phase (red) in Event~1 (a, b) and Event~2 (c, d).}
\end{figure}

It was noticed previously that an enhancement of the broadband electrostatic noise was detected during the recovery phase in Event~1 \cite{runov25b}. To characterise the wave power level during pre-storm and recovery phase interval, we plotted integral wave power from EFI and SCM FBK five frequency bands in the range from 9.05 to 2689.0\,Hz (Figure~\ref{artm_FBK}). Evidently, the electrostatic wave power increased by an order of a magnitude during the recovery phase compare to the pre-storm level.

\begin{figure}[t]
\includegraphics[width=\textwidth]{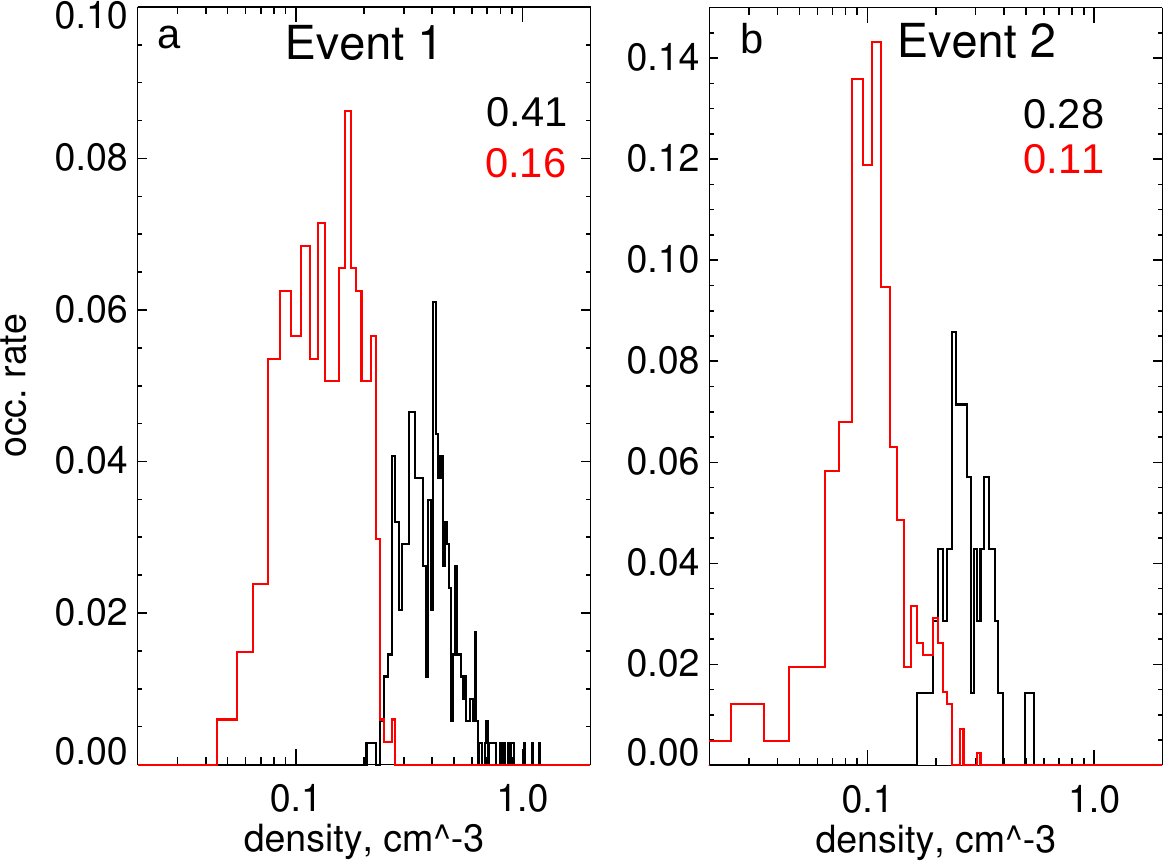}
\caption{\label{artm_ni}
Particle density distributions during the pre-storm interval (black) and the recovery phase (red) in Event~1 (a) and Event~2 (b). The corresponding median values (in cm$^{-3}$) are printed in corresponding colors.
}
\end{figure}

Figure \ref{artm_ni} shows distributions of density (ion density was used to avoid the uncertainty caused by the attracted photo-electrons) observed during pre-storm interval (black) and recovery phase in Event~1 (panel a) and Event~2 (panel b). In both events the recovery phase density was by a factor of 0.6 lower than pre-storm plasma sheet density.

\begin{figure}[t]
\includegraphics[width=\textwidth]{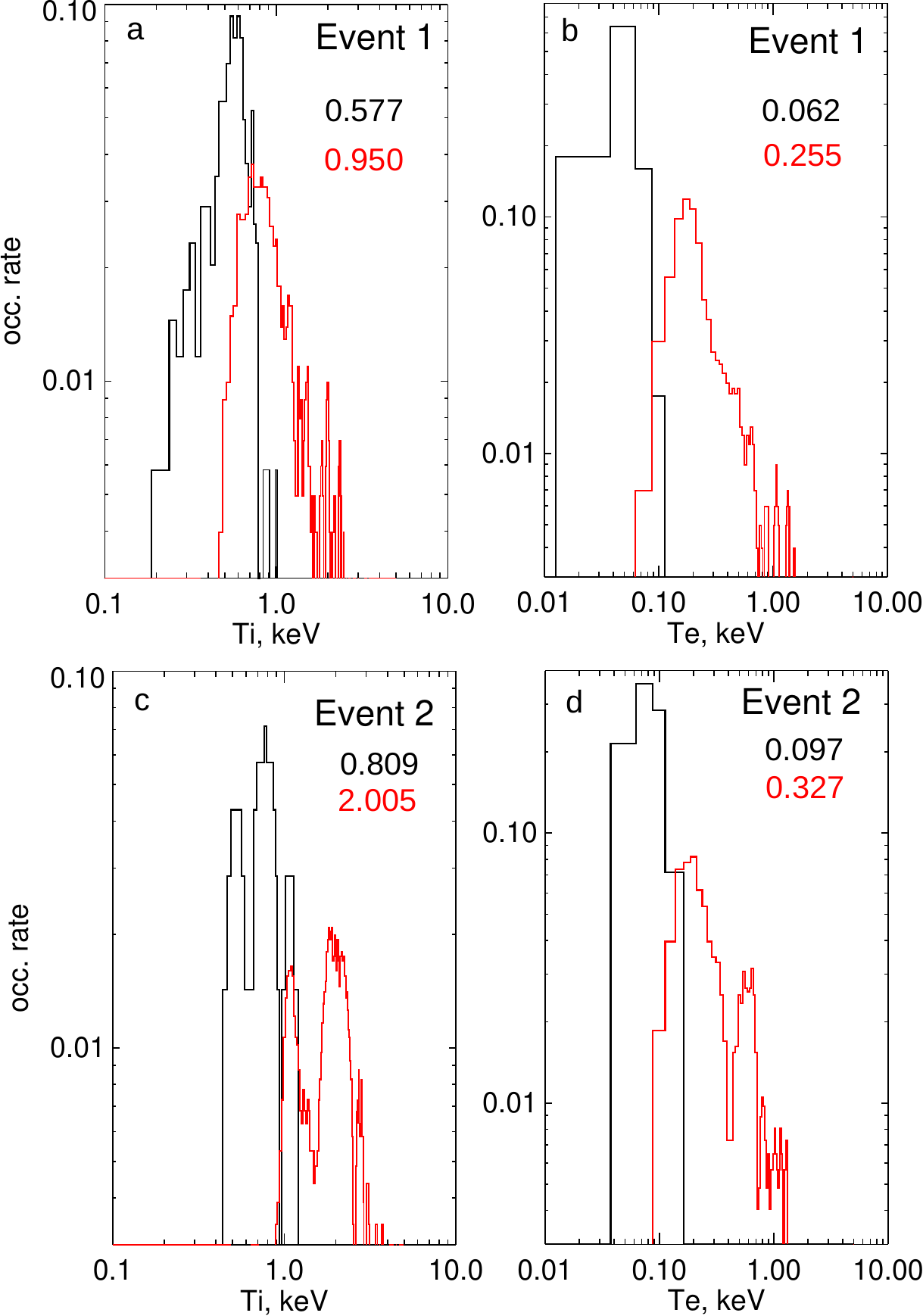}
\caption{\label{artm_Tie}
Ion (a, c) and electron (b, d)  temperatures distributions during the pre-storm interval (black) and the recovery phase (red) in Event~1 (a, b) and Event~2 (c, d). The corresponding median values (in keV) are printed in corresponding colors.
}
\end{figure}

\begin{figure}[t]
\includegraphics[width=\textwidth]{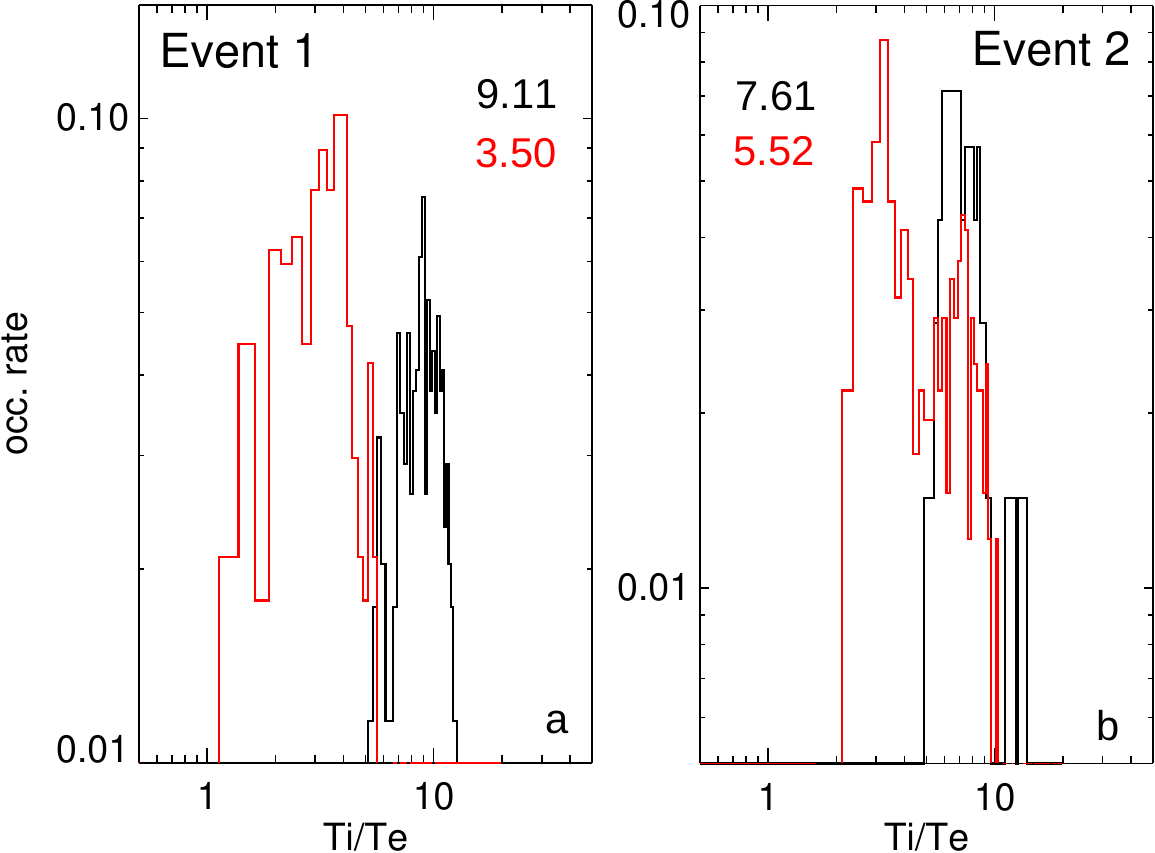}
\caption{\label{artm_Tier}
Ion to electron temperature ratio ($T_i/T_e$) distributions during the pre-storm interval (black) and the recovery phase (red) in Event~1 (a) and Event~2 (b). The corresponding median values are printed in corresponding colors.
}
\end{figure}

Figure \ref{artm_Tie} shows distributions of ion and electron temperatures during pre-storm and recovery phase intervals for Event~1 and 2. Evidently, in both events ion and electron temperatures are higher on the storm recovery phases compare to the pre-storm intervals. The relative electron temperature gain
$\delta{T_e}=\langle{{T_e}_{rec}}\rangle-\langle{{T_e}_{pre}}\rangle/\langle{{T_e}_{pre}}\rangle$,
however, larger than the relative ion temperature gain
$\delta{T_i}=\langle{{T_i}_{rec}}\rangle-\langle{{T_i}_{pre}}\rangle/\langle{{T_i}_{pre}}\rangle$, where indices $pre$ and $rec$ indicate pre-storm and recovery phase intervals, respectively. In Event~1 $\delta{T_i}$=0.65, whereas $\delta{T_e}$=3.13. The difference between $\delta{T_i}$ and $\delta{T_e}$ was smaller: $\delta{T_i}$=1.5 versus
$\delta{T_e}$=2.6.

The preferential electron energization with respect to that of ions is more evident in the ion to electron temperature ratio ($T_i/T_e$) distributions shown in Figure~\ref{artm_Tier}. Again, the effect is more pronounced for Event~1. It may be because the ion temperature was lower during pre-storm interval in Event~1 compare to Event~2. It is also notable, that in both events both $T_i$ and $T_e$ distributions exhibit a double peak shape, which is more pronounced in Event~2 distributions.

\begin{figure}[t]
\includegraphics[width=\textwidth]{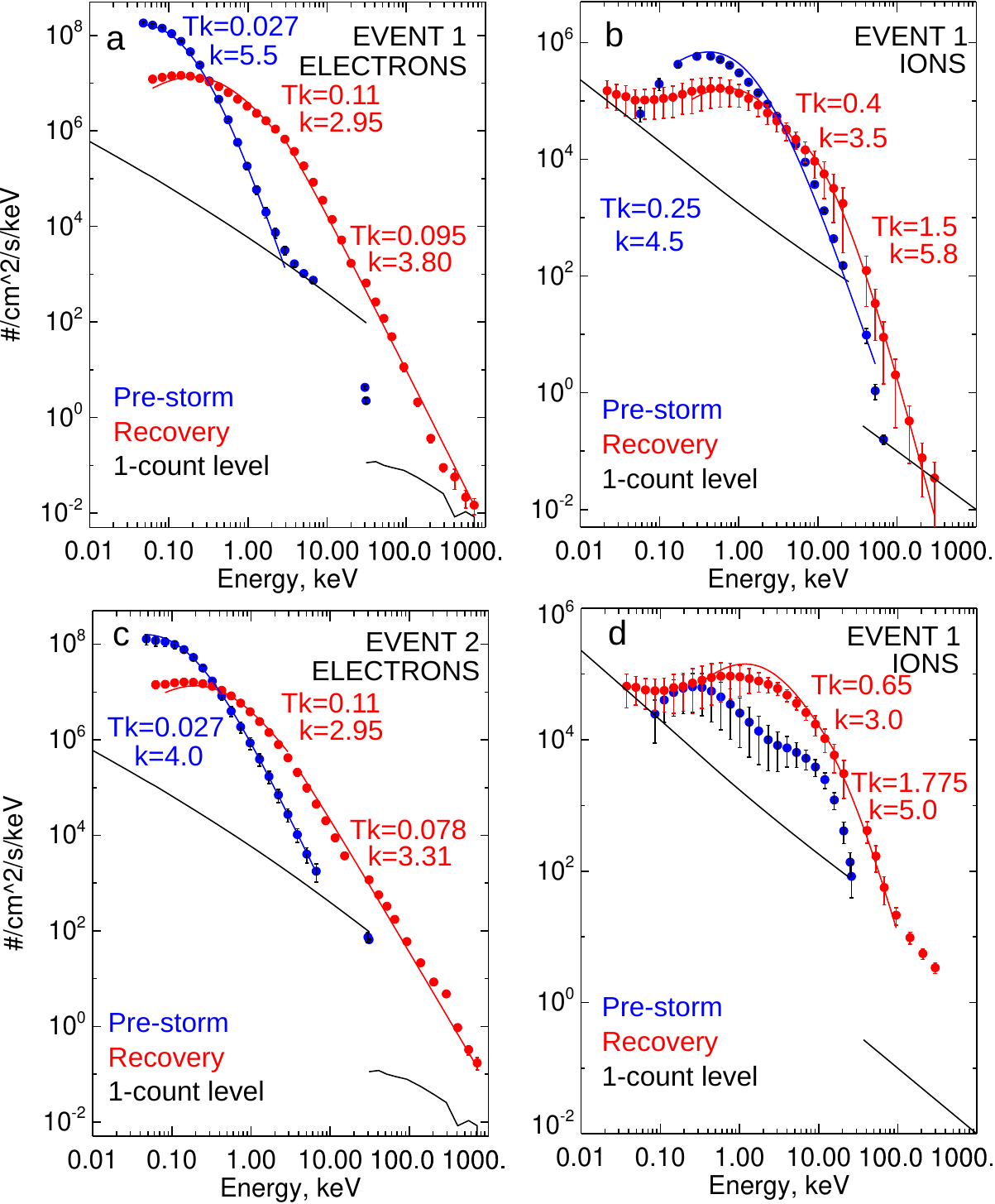}
\caption{\label{artm_specs}
Electron (a, c) and ion (b, d) particle spectra collected during the pre-storm interval (blue dots) and the recovery phase (red dots) in Event~1 (a, b) and Event~2 (c, d), respectively. The error-bars indicate the standard deviation ${\sigma}/\sqrt{N}$, where $N$ is the number of points.
The corresponding curves shows the fit by the Kappa function. Note that thermal parts and energy tails of recovery phase spectra are fitted by separately. The corresponding fitting parameters (spectral temperature $T_k)$ and kappa exponent $\kappa$) are shown in plots.
Black curves indicate 1-count levels for ESA and SST instruments.
}
\end{figure}

Figure \ref{artm_specs} shows electron and ion particle spectra (particle number flux in $\#/cm^2/s/str/keV$ vs energy in $keV$) observed during pre-storm interval and on the recovery phase in Event~1 and 2. Evidently, the recovery phase electron spectra in both Event~1 and 2 changed dramatically with respect to those at pre-storm interval. The fluxes at energy range from $\sim$10 to 700\,keV were not present at the pre-storm interval. During the recovery phase electron fluxes in this range were observed robustly. At the energy of 100\,keV the recovery phase electron flux exceed the 1-count level by two orders of magnitude. A caution is needed in Event~2, which was during the SEP event. Despite our efforts to clean up energy-independent background, some remnant fluxes may still be present. The functional form of the spectrum, however, suggest physical nature of the observed fluxes. The recovery phase electron spectra are harder (the $\kappa$-exponent $3<{\kappa}<4$  than pre-storm ones ($4<{\kappa}<5$). At the same time, the spectra do not indicate an injection of a new, accelerated population and consistent rather with heating of the pre-existed population: fluxes at energies lower that $\approx$0.5\,keV became lower, whereas fluxes at higher energies increased.

Ion spectra, obtained during the recovery phase, also show substantial changes, however, not nearly that dramatic as electron ones. In Event~1, ion fluxes, similar to electron ones, indicate heating of the pre-existed population. In Even~2, however, the spectra obtained prior to the storm and on the recovery phase are different. A significant variations, indicated by error-bars, are also notable. A flattening of the Event~2 recovery phase ion spectrum is, most likely, due to instrumental effect: penetration of energetic electrons through instrument shielding and the magnetic deflector.

\section{Summary and Discussion}
The presented observations revealed preferential and strong energization of electrons - up to relativistic energies - during the storm's recovery phase. In both events, the differential electron fluxes in the energy range of 10 to 100\,keV, barely present prior to storm, increased up to 10$^5$ and 10$/cm^2/s/sr/keV$, respectively. Integral fluxes at energies $E>100$\,keV, not present prior to storm, increased at the recovery phase to 10$/cm^2/s$. The average electron temperature in both events increased at the recovery phase by a factor of 4.
Energization of ions, although present, was not nearly that strong as for electrons. The average ion temperature increase at recovery phase with respect to that at pre-storm interval by factors 1.6 in Event~1 and 2.4 in Event~2.

Another potentially important observation is the increase in the electric field and in the electrostatic fluctuation level. More pronounced in Event~1, the equatorial electric field $E_{eq}=(E_x + E_y)^{1/2}$ increased during the recovery phase to
$\approx$2\,mV/m. Moreover, this electric field was generated by a kinetic process: neither $|{\bf v}_i \times {\bf B}|$ nor ${\bf v}_e \times {\bf B}$ exceeded a level of $\approx$1\,mV/m. In both events, the integral power of electrostatic fluctuations increased at recovery phase to $\approx$1\,mV/m versus $\approx$0.1\,mV/m during the pre-storm interval.

What are possible source(s) and mechanism(s) of the observed electron energization up to relativistic energies in the magnetotail?

The very first possibility is the solar energetic particle (SEP) penetration into the magnetosphere. This hypothesis, however, is not supported by the observations during Event~1. It has been shown, SEP was not observed neither by WIND in the solar wind, nor by ARTEMIS in the magnetotail \cite{runov25b}. It is a possibility for Event~2, though. Indeed, SEP-related background energy-independent fluxes are evident in the SST energy-time spectrograms during Event~2. Because SEP energy is higher than the SST range, the SEP particles hit all the detectors in the SST stack, which produce an energy-independent background in counts. After the background removal, the energy dependent counts emerged, indicating the other population, which was energized, likely, independently of SEP.

In Event~1 the inner THEMIS probes encountered the magnetosheath during the pre-storm interval and the storm main and recovery phases (Fig.~\ref{thasplot}. No fluxes of electrons at energy $E>$30\,keV (SST range) were detected during the magnetosheath excursions. Thus, at least for Event~1, the bow shock acceleration mechanisms may be ruled out. We cannot exclude this possibility for Event~2, however, because of lack of simultaneous magnatosheath observations.

Can the relativistic electrons, detected by ARTEMIS, escape from the outer radiation belt into the magnetosheath or PSBL and reenter the nightside magnetosphere from the mid-tail magntosheath or the plasma sheet from PSBL? Possible, but questionable.
The reentry through PSBL is not supported by NOAA observations in Event~1 see Fig.~4 in \citeA{runov25b}. In this case, the sun-synchronous NOAA-15 spacecraft detected clear drop in the energetic electron fluxes when exit the radiation belt and, later on, sudden reappearance of the fluxes at larger geomagnetic latitudes (larger geocentric distances). These observations cannot rule out a possibility for relativistic electrons leaking from the outer radiation belt to
reentry through the mid-magnetotail flank as suggested in
\cite{sibeck85, runov25b}.

Can the electrons be energized to relativistic energies via multiple bouncing in the near-Earth plasma sheet \cite<e.g.,>{runov25a} and retreat subsequently to mid-distant tail? This mechanism requires a stable trapping of the energetic electrons within tailward stretching magnetic flux tubes. It is hardly possible, because, due to magnetic field curvature radius decreasing this electrons will be subject to azimuthal drift and will be lost from the stretching flux tube. A rough estimation yields with the maximum tail-along path of $\sim$100\,keV electrons of 4\,R$_E$. Moreover, initially field-align electrons would be quickly scattered by the magnetic field curvature, demagnetized, and, again, moved azimuthally by the dawn-dusk electric field away form the plasma sheet.

The most intriguing question is may electrons be energized to relativistic energies locally in the mid-distant magnetotail? Magnetic reconnection is the strongest mechanism of magnetic energy conversion into the particle kinetic energy. However, the lobe magnetic field strength was $B_L\approx$10\,nT at $R\sim$60\,R$_E$ on the storm recovery phase (Fig.~\ref{bl_beq}). It does not provide enough energy per particle for energization to $E\geq$100\,keV. It is possible, however, that the energy conversion in such conditions (weak and turbulent magnetic field) operate continuously via multiple, sporadic electron-only reconnection, similar to that in the magnetosheath \cite{phan18}. Observations of electron-only reconnection in the magnetotail were also reported
\cite<e.g.,>{sanlu20, hubbert21}. The electron-only reconnection paradigm might also explain the preferential energization of electrons over ions. The observed electron inflow (Fig.~\ref{artm_viez}) supports the electron-only reconnection scenario. Question is whether electron-only reconnection is powerful enough to energize electrons to 100s of keV. Kinetic simulations are required to address this question.

Another possibility is stochastic acceleration in a highly fluctuating electric field \cite{ergun18, usanova22}. Effective electron energization and pitch-angle scattering can occur if the correlation length scale of electric field fluctuations is comparable to or smaller than the electron gyroradius \cite<e.g.>[and references therein]{oka23SSR}. The observed increase in electrostatic fluctuation power (Fig.~\ref{artm_FBK}) supports this suggestion.

Turbulence and electron-only reconnection are highly coupled \cite<e.g.,>{stawarz24SSR}. Thus the two mechanisms are not mutually exclusive. Whether they are powerful enough to explain the observed fluxes of relativistic electrons in the near-Moon magnetotail remains an open question.

\section{Conclusions}
We studied two storm magnetic storm events during which the lunar-orbiting ARTEMIS probes traversed the magnetitail. In both events enhancements of electron fluxes at relativistic energies $E\geq$100\,keV were observed on the storms' recovery phase. Our analysis suggests that these electrons were energized to relativistic energies internally in the plasma sheet. The acceleration process was, likely, associated with enhancement in broadband electrostatic fluctuation power on the recovery phase. Preferential energization of electrons, compare to ions, may suggest that electron-only reconnection might be in action. Further analysis, including kinetic simulations is required to clarify this hypothesis.

%
%

\section*{Open Research Section}
The THEMIS data used in our study are available at UC Berkeley repository at
(\url{http://themis.ssl.berkeley.edu/index.shtml}). The OMNI data are available via the SPEDAS system (\url{http://spedas.org}).

\acknowledgments
This work was supported by the  NASA contract NAS5-02099, NASA grants
80NSSC21K1407 and 80NSSC22K1929. AR thanks ISSI and memebers of International Team led by N.~Ganjushkina and M.~Liemohn for support and fruitful discussions.
We acknowledge D. Larson and R. P.~Lin for use of SST data; C.W.~Carlson and J.P.~McFadden for use of ESA data;  FGM data provided under the lead of the Technical University of Braunschweig and with financial support through the German Ministry for Economy and Technology and the German Center for Aviation and Space (DLR) under contract 50 OC 0302.
We thank J.~Lewis for help with the SPEDAS software.

%
%

\bibliography{refdb}

@ARTICLE{angelopoulos08a,
   author = {V. Angelopoulos and D. Sibeck and C. W. Carlson and J. P. {McFadden} and D. Larson and R. P. Lin and J. W. Bonnell and F. S. Mozer and R. Ergun and C. Cully and K. H. Glassmeier and U. Auster and A. Roux and O. {LeContel} and S. Frey and T. Phan and S. Mende and H. Frey and E. Donovan and C. T. Russell and R. Strangeway and J. Liu and I. Mann and J. Rae and J. Raeder and X. Li and W. Liu and H. J. Singer and V. A. Sergeev and S. Apatenkov and G. Parks and M. Fillingim and J. Sigwarth },
    title = {First Results from the {THEMIS} mission},
  journal = {Space Sci. Rev},
     year = 2008,
   volume = {141},
      pages = {453-476} }

@ARTICLE{angelopoulos11,
   author = {V. Angelopoulos},
    title = {The {ARTEMIS} Mission},
  journal = {Space Sci. Rev},
     year = 2011,
   volume = {165},
      pages = {3-25, doi:10.1007/s11214-010-9687-2} }

@ARTICLE{angelopoulos19,
   author = {V.~Angelopoulos and A.~V.~Artemyev and T.~D.~Phan and Y.~Miyashita}, 
    title = {Near-Earth Magnetotail Reconnection Powers Space Storms},
  journal = {Nature Physics},
     year = 2019,
   volume = {16},
      pages = { doi:10.1038/s41567-019-0749-4} }

@article{auster08,
    author = {H. U. Auster and K. H. Glassmeier and W. Magnes and O. Aydogar and W. Baumjohann and D. Constantinescu and D. Fischer and K. H. Fornacon and E. Georgescu and P. Harvey and O. Hillenmaier and
              R. Kroth and M. Ludlam and Y. Narita and R. Nakamura and K. Okrafka and F. Plaschke and I. Richter and H. Schwarzl and B. Stoll and A. Valavanoglou and M. Wiedemann},
    title = {The {THEMIS} Fluxgate Magnetometer},
    journal = {Space Sci. Rev.},
    year = 2008,
    volume=141,
    pages = {235-264}}

@ARTICLE{baker77,
   author = {D. N. Baker and E. C. Stone},
    title = {Observations of energetic electrons {E} no less than about 200 keV in the {Earth's} magnetotail: Plasma sheet and fireball observations},
  journal = {J. Geophys. Res.},
     year = 1977,
   volume = 82,
    pages = {1532-1546} }

@article{birn14,
    author = {J. Birn and A.~Runov and M.~Hesse}, 
    title = {Energetic Electrons in Dipolarization Events: Spatial Properties
               and Anisotropy}, 
    journal = {J. Geophys. Res.}, 
    year = 2014,    
    volume = 119, 
    pages = { 3604-3616}, 
    doi={10.1002/2013JA019738}}

@article{bonnell08,
    author = {J. W. Bonnell and F. S. Mozer and G. T. Delory and A. J. Hull and R. E. Ergun and C. M. Cully and V. Angelopoulos and P. R. Harvey},
    title = {The Electric Field Instrument ({EFI}) for {THEMIS}},
    journal = {Space Sci. Rev.},
    year = 2008,
    volume = 141,
    pages = {303-341}}

@ARTICLE{cohen17,
       author = {{Cohen}, Ian J. and {Mauk}, Barry H. and {Anderson}, Brian J. and {Westlake}, Joseph H. and {Sibeck}, David G. and {Turner}, Drew L. and {Fennell}, Joseph F. and {Blake}, J. Bern and {Jaynes}, Allison N. and {Leonard}, Trevor W. and {Baker}, Daniel N. and {Spence}, Harlan E. and {Reeves}, Geoff D. and {Giles}, Barbara J. and {Strangeway}, Robert J. and {Torbert}, Roy B. and {Burch}, James L.},
        title = "{Statistical analysis of MMS observations of energetic electron escape observed at/beyond the dayside magnetopause}",
      journal = {Journal of Geophysical Research (Space Physics)},
         year = 2017,
        month = sep,
       volume = {122},
       number = {9},
        pages = {9440-9463},
          doi = {10.1002/2017JA024401},
       adsurl = {https://ui.adsabs.harvard.edu/abs/2017JGRA..122.9440C}
}

@ARTICLE{cohen21,
       author = {{Cohen}, I.~J. and {Turner}, D.~L. and {Michael}, A.~T. and {Sorathia}, K.~A. and {Ukhorskiy}, A.~Y.},
        title = "{Investigating the Link Between Outer Radiation Belt Losses and Energetic Electron Escape at the Magnetopause: A Case Study Using Multi-Mission Observations and Simulations}",
      journal = {Journal of Geophysical Research (Space Physics)},
         year = 2021,
        month = jun,
       volume = {126},
       number = {6},
          eid = {e29261},
        pages = {e29261},
          doi = {10.1029/2021JA029261},
       adsurl = {https://ui.adsabs.harvard.edu/abs/2021JGRA..12629261C}
}

@ARTICLE{delcourt99,
       author = {{Delcourt}, D.~C. and {Sauvaud}, J.-A.},
        title = "{Populating of the cusp and boundary layers by energetic (hundreds of keV) equatorial particles}",
      journal = {J. Geophys. Res.},
         year = 1999,
        month = oct,
       volume = {104},
       number = {A10},
        pages = {22635-22648},
          doi = {10.1029/1999JA900251},
       adsurl = {https://ui.adsabs.harvard.edu/abs/1999JGR...10422635D}
}

@ARTICLE{ergun18,
       author = {{Ergun}, R.~E. and {Goodrich}, K.~A. and {Wilder}, F.~D. and {Ahmadi}, N. and {Holmes}, J.~C. and {Eriksson}, S. and {Stawarz}, J.~E. and {Nakamura}, R. and {Genestreti}, K.~J. and {Hesse}, M. and {Burch}, J.~L. and {Torbert}, R.~B. and {Phan}, T.~D. and {Schwartz}, S.~J. and {Eastwood}, J.~P. and {Strangeway}, R.~J. and {Le Contel}, O. and {Russell}, C.~T. and {Argall}, M.~R. and {Lindqvist}, P. -A. and {Chen}, L.~J. and {Cassak}, P.~A. and {Giles}, B.~L. and {Dorelli}, J.~C. and {Gershman}, D. and {Leonard}, T.~W. and {Lavraud}, B. and {Retino}, A. and {Matthaeus}, W. and {Vaivads}, A.},
        title = "{Magnetic Reconnection, Turbulence, and Particle Acceleration: Observations in the Earth's Magnetotail}",
      journal = {Geophys.~Res.~Lett.},
         year = 2018,
        month = apr,
       volume = {45},
       number = {8},
        pages = {3338-3347},
          doi = {10.1002/2018GL076993},
       adsurl = {https://ui.adsabs.harvard.edu/abs/2018GeoRL..45.3338E}
}

@article{fujimoto01,
    author = {M. Fujimoto and T. Nagai and N. Yokokawa and Y. Yamade and T. Mukai and Y. Saito and S. Kokubun},
    title = {Tailward electrons at the lobe-plasma sheet interface detected upon dipolarizations},
    journal = {J. Geophys. Res.},
    year = 2001,
    volume = 106,
    pages = {21255-21262} }

@ARTICLE{gabrielse14,
   author = {C. {Gabrielse} and V. {Angelopoulos} and A. {Runov} and D. Turner},
    title = {Statistical
characteristics of particle injections throughout the equatorial magnetotail},
  journal = {J. Geophys. Res.},
     year = 2014,
   volume = 119,
    pages = {2512-2535}, 
    doi={10.1002/2013JA019638} }

@ARTICLE{gabrielse16,
       author = {{Gabrielse}, Christine and {Harris}, Camilla and {Angelopoulos}, Vassilis and {Artemyev}, Anton and {Runov}, Andrei},
        title = {The role of localized inductive electric fields in electron injections around dipolarizing flux bundles},
      journal = {Journal of Geophysical Research (Space Physics)},
         year = "2016",
        month = "Oct",
       volume = {121},
       number = {10},
        pages = {9560-9585},
          doi = {10.1002/2016JA023061},
       adsurl = {https://ui.adsabs.harvard.edu/abs/2016JGRA..121.9560G}
}

@ARTICLE{gabrielse22SW,
       author = {{Gabrielse}, Christine and {Lee}, Justin H. and {Claudepierre}, Seth and {Walker}, Don and {O'Brien}, Paul and {Roeder}, James and {Lao}, Yao and {Grovogui}, Jann and {Turner}, Drew L. and {Runov}, Andrei and {Boyd}, Alexander and {Fennell}, Joseph and {Blake}, J. Bernard and {Lopez}, Kevin and {Miyoshi}, Yoshizumi and {Keika}, Kunihiro and {Higashio}, Nana and {Shinohara}, Iku and {Imajo}, Shun and {Kurita}, Satoshi and {Mitani}, Takefumi},
        title = {Radiation Belt Daily Average Electron Flux Model (RB-Daily-E) From the Seven-Year Van Allen Probes Mission and Its Application to Interpret GPS On-Orbit Solar Array Degradation},
      journal = {Space Weather},
         year = 2022,
        month = nov,
       volume = {20},
       number = {11},
          eid = {e2022SW003183},
        pages = {e2022SW003183},
          doi = {10.1029/2022SW003183},
       adsurl = {https://ui.adsabs.harvard.edu/abs/2022SpWea..2003183G}
}

@article{hubbert21,
author = {Hubbert, M. and Qi, Y. and Russell, C. T. and Burch, J. L. and Giles, B. L. and Moore, T. E.},
title = {Electron-Only Tail Current Sheets and Their Temporal Evolution},
journal = {Geophysical Research Letters},
volume = {48},
number = {5},
pages = {e2020GL091364},
doi = {https://doi.org/10.1029/2020GL091364},
url = {https://agupubs.onlinelibrary.wiley.com/doi/abs/10.1029/2020GL091364},
eprint = {https://agupubs.onlinelibrary.wiley.com/doi/pdf/10.1029/2020GL091364},
note = {e2020GL091364 2020GL091364},
year = {2021}
}

@ARTICLE{kamaletdinov24,
       author = {{Kamaletdinov}, S.~R. and {Artemyev}, A.~V. and {Runov}, A. and {Angelopoulos}, V.},
        title = "{Characteristics of Thin Magnetotail Current Sheet Plasmas at Lunar Distances}",
      journal = {Journal of Geophysical Research (Space Physics)},
         year = 2024,
        month = aug,
       volume = {129},
       number = {8},
          eid = {e2024JA032755},
        pages = {e2024JA032755},
          doi = {10.1029/2024JA032755},
       adsurl = {https://ui.adsabs.harvard.edu/abs/2024JGRA..12932755K}
}

@ARTICLE{kamaletdinov25GRL,
       author = {{Kamaletdinov}, S.~R. and {Artemyev}, A.~V. and {Runov}, A. and {Angelopoulos}, V.},
        title = "{Ion Kinetics in Thin Current Sheets at Lunar Distances}",
      journal = {Geophys.~Res.~Lett.},
         year = 2025,
        month = may,
       volume = {52},
       number = {9},
          eid = {e2024GL114522},
        pages = {e2024GL114522},
          doi = {10.1029/2024GL114522},
       adsurl = {https://ui.adsabs.harvard.edu/abs/2025GeoRL..5214522K}
}

@ARTICLE{liu22ApJ,
       author = {{Liu}, Zixuan and {Wang}, Linghua and {Guo}, Xinnian},
        title = "{Acceleration of Solar Wind Suprathermal Electrons at the Earth's Bow Shock}",
      journal = {The~Astrophysical~Journal},
         year = 2022,
        month = aug,
       volume = {935},
       number = {1},
          eid = {39},
        pages = {39},
          doi = {10.3847/1538-4357/ac8157},
       adsurl = {https://ui.adsabs.harvard.edu/abs/2022ApJ...935...39L}
}

@ARTICLE{sanlu20,
       author = {{Lu}, San and {Wang}, Rongsheng and {Lu}, Quanming et al.},
        title = "{Magnetotail reconnection onset caused by electron kinetics with a strong external driver}",
      journal = {Nature Communications},
         year = 2020,
        month = oct,
       volume = {11},
          eid = {5049},
        pages = {5049},
          doi = {10.1038/s41467-020-18787-w},
       adsurl = {https://ui.adsabs.harvard.edu/abs/2020NatCo..11.5049L}
}

@article{mcfadden08,
    author = {J. P. {McFadden} and C. W. Carlson and D. Larson and V. Angelopolos and M. Ludlam and R. Abiad and B. Elliot},
    title = {The {THEMIS} {ESA} Plasma Instrument and In-flight Calibration},
    journal = {Space Sci. Rev.},
    year = 2008,
    volume = {141},
    pages = {277-302} }

@ARTICLE{oka06,
       author = {{Oka}, M. and {Terasawa}, T. and {Seki}, Y. and {Fujimoto}, M. and {Kasaba}, Y. and {Kojima}, H. and {Shinohara}, I. and {Matsui}, H. and {Matsumoto}, H. and {Saito}, Y. and {Mukai}, T.},
        title = "{Whistler critical Mach number and electron acceleration at the bow shock: Geotail observation}",
      journal = {Geophys.~Res.~Lett},
         year = 2006,
        month = dec,
       volume = {33},
       number = {24},
          eid = {L24104},
        pages = {L24104},
          doi = {10.1029/2006GL028156},
       adsurl = {https://ui.adsabs.harvard.edu/abs/2006GeoRL..3324104O}
}

@ARTICLE{oka22POP,
       author = {{Oka}, Mitsuo and {Phan}, Tai and {{\O}ieroset}, Marit and {Turner}, Drew and {Drake}, James and {Li}, Xiaocan and {Fuselier}, Stephen and {Gershman}, Daniel and {Giles}, Barbara and {Ergun}, Robert and {Torbert}, Roy and {Wei}, Hanying and {Strangeway}, Robert and {Russell}, Christopher and {Burch}, James},
        title = "{Electron energization and thermal to non- thermal energy partition during earth's magnetotail reconnection}",
      journal = {Physics of Plasmas},
         year = 2022,
        month = may,
       volume = {29},
       number = {5},
          eid = {052904},
        pages = {052904},
          doi = {10.1063/5.0085647},
       adsurl = {https://ui.adsabs.harvard.edu/abs/2022PhPl...29e2904O}
}

@ARTICLE{oka23SSR,
       author = {{Oka}, Mitsuo and {Birn}, Joachim and {Egedal}, Jan and {Guo}, Fan and {Ergun}, Robert E. and {Turner}, Drew L. and {Khotyaintsev}, Yuri and {Hwang}, Kyoung-Joo and {Cohen}, Ian J. and {Drake}, James F.},
        title = "{Particle Acceleration by Magnetic Reconnection in Geospace}",
      journal = {Space~Sci.~Rev.},
         year = 2023,
        month = dec,
       volume = {219},
       number = {8},
          eid = {75},
        pages = {75},
          doi = {10.1007/s11214-023-01011-8},
archivePrefix = {arXiv},
       eprint = {2307.01376},
 primaryClass = {physics.space-ph},
       adsurl = {https://ui.adsabs.harvard.edu/abs/2023SSRv..219...75O}
}

@ARTICLE{oieroset02,
       author = {{{\O}ieroset}, M. and {Lin}, R.~P. and {Phan}, T.~D. and {Larson}, D.~E. and {Bale}, S.~D.},
        title = "{Evidence for Electron Acceleration up to \raisebox{-0.5ex}\textasciitilde300 keV in the Magnetic Reconnection Diffusion Region of Earth's Magnetotail}",
      journal = {Phys.~Rev.~Lett.},
         year = 2002,
        month = oct,
       volume = {89},
       number = {19},
          eid = {195001},
        pages = {195001},
          doi = {10.1103/PhysRevLett.89.195001},
       adsurl = {https://ui.adsabs.harvard.edu/abs/2002PhRvL..89s5001O}
}

@ARTICLE{phan18,
       author = {{Phan}, T.~D. and {Eastwood}, J.~P. and {Shay}, et al.},
        title = "{Electron magnetic reconnection without ion coupling in Earth's turbulent magnetosheath}",
      journal = {Nature},
         year = 2018,
        month = may,
       volume = {557},
       number = {7704},
        pages = {202-206},
          doi = {10.1038/s41586-018-0091-5},
       adsurl = {https://ui.adsabs.harvard.edu/abs/2018Natur.557..202P}
}

@article{richardson96,
    author = {I. G. Richardson and C. J. Owen and J. A. Slavin},
    title = {Energetic ($>$0.2 {MeV}) electron bursts in the deep geomagnetic tail observed by the {Goddard Space Flight Center} experiment on {ISEE 3}: Association with geomagnetic substorms},
    journal = {J. Geophys. Res.},
    year = 1996,
    volume = 101,
    pages = {2723-2740} }

@article{roux08,
author = {A. Roux and O. {Le Contel} and C. Coillot and A. Bouabdellah and B. {de la Porte} and D. Alison and S. Ruocco and M. C. Vassal},
title = {The Search Coil Magnetometer for THEMIS},
journal = {Space Sci. Rev.},
year = 2008,
volume = {141},
pages = {	265-275}
}

@article{runov15,
    author = {A. Runov and V. Angelopoulos and C.~Gabrielse and J.~Liu and
D.~L.~Turner and X.-Z.~Zhou},
    title = {Average Thermodynamic and Spectral Properties of Plasma in and
Around Dipolarizing Flux Bundles},
    journal = {J. Geophys. Res.},
    year = 2015,
    volume = 120,
    doi= {10.1002/2015JA021166} }

@article{runov18,
    author = {A. Runov and V.~Angelopoulos and A.~V.~Artemyev and S.~Lu and X.-Z. Zhou},
    title = {Near-Earth Reconnection Ejecta at Lunar Distances},
    journal = {J. Geophys. Res.},
    year = 2018,
    volume = 123,
    pages={2736-2744},
    doi= {10.1002/2017JA025079} }

@ARTICLE{runov23,
       author = {{Runov}, A. and {Angelopoulos}, V. and {Khurana}, K. and {Liu}, J. and {Balikhin}, M. and {Artemyev}, A.~V.},
        title = "{Properties of Quiet Magnetotail Plasma Sheet at Lunar Distances}",
      journal = {Journal of Geophysical Research (Space Physics)},
         year = 2023,
        month = nov,
       volume = {128},
       number = {11},
          eid = {e2023JA031908},
        pages = {e2023JA031908},
          doi = {10.1029/2023JA031908},
       adsurl = {https://ui.adsabs.harvard.edu/abs/2023JGRA..12831908R}
}

@ARTICLE{runov25a,
       author = {{Runov}, A. and {Angelopoulos}, V. and {Artemyev}, A.~V. and {Birn}, J. and {Engebretson}, M.~J. and {Weygand}, J.~M. and {Xu}, Z.},
        title = "{THEMIS Observations of Relativistic Electrons at the Nightside Transition Region During HILDCAA Events}",
      journal = {Journal of Geophysical Research (Space Physics)},
         year = 2025,
        month = feb,
       volume = {130},
       number = {2},
        pages = {2024JA033179},
          doi = {10.1029/2024JA033179},
       adsurl = {https://ui.adsabs.harvard.edu/abs/2025JGRA..13033179R}
}

@ARTICLE{runov25b,
       author = {{Runov}, A. and {Angelopoulos}, V. and {Artemyev}, A.~V. and {Shi}, X. and {Gabrielse}, C.},
        title = "{Prolonged Intervals of Relativistic Electron Storm-Time Flux Enhancements in the Magnetotail at Lunar Distance}",
      journal = {Geophys.~Res.~Lett},
         year = 2025,
        month = aug,
       volume = {52},
       number = {16},
          eid = {e2025GL116847},
        pages = {e2025GL116847},
          doi = {10.1029/2025GL116847},
       adsurl = {https://ui.adsabs.harvard.edu/abs/2025GeoRL..5216847R}
}

@ARTICLE{sarafopoulos04,
   author = {D.~V. {Sarafopoulos}  and E.~T. {Sarris} and V.~ Lutsenko},
    title = {Evidence for energetic electron and ion dispersive microinjections in the {Earth's} magnetotail as far as 27\,R$_E$},
  journal = {Ann.~Geo},
     year = 2004,
   volume = 22,
    pages = {527-535} }

@ARTICLE{sarris76,
   author = {E. T. Sarris and S. M. Krimigis and T. P. Armstrong},
    title = {Observations of magnetospheric bursts of high-energy protons and electrons at approximately 35 {Earth} radii with {Imp 7}},
  journal = {J. Geophys. Res.},
     year = 1976,
   volume = 81,
    pages = {2341-2355} }

@ARTICLE{sarris96,
   author = {E. T. Sarris and V. Angelopoulos and R.~W. {McEntire} and D.~J. Williams and S.~M. Krimigis and A.~T.~Y. Lui and E.~C. Roelof and S. Kokubun},
    title = {Detailed observations of a burst of energetic particles in the deep magnetotail by {Geotail}},
  journal = {J. Geomag. Geoelectr.},
     year = 1996,
   volume = 48,
    pages = {649-656} }

@article{schodel01a,
    author = {R. Sch{\"o}del and W. Baumjohann and R. Nakamura and V. A. Sergeev and T. Mukai},
    title = {Rapid flux transport in the central plasma sheet},
    journal = {J. Geophys. Res.},
    year = 2001,
    volume = 106,
    pages = {301-313} }

@ARTICLE{sibeck85,
   author = {{Sibeck}, D.~G. and {Siscoe}, G.~L. and {Slavin}, J.~A. and
	{Smith}, E.~J. and {Tsurutani}, B.~T.},
    title = {Magnetic field properties of the distant magnetotail magnetopause and boundary layer},
  journal = {J. Geophys. Res.},
     year = 1985,
    month = oct,
   volume = 90,
    pages = {9561-9575},
      doi = {10.1029/JA090iA10p09561},
   adsurl = {http://adsabs.harvard.edu/abs/1985JGR....90.9561S}
}

@ARTICLE{sorathia17,
       author = {{Sorathia}, K.~A. and {Merkin}, V.~G. and {Ukhorskiy}, A.~Y. and {Mauk}, B.~H. and {Sibeck}, D.~G.},
        title = "{Energetic particle loss through the magnetopause: A combined global MHD and test-particle study}",
      journal = {Journal of Geophysical Research (Space Physics)},
         year = 2017,
        month = sep,
       volume = {122},
       number = {9},
        pages = {9329-9343},
          doi = {10.1002/2017JA024268},
       adsurl = {https://ui.adsabs.harvard.edu/abs/2017JGRA..122.9329S}
}

@ARTICLE{sorathia18,
       author = {{Sorathia}, Kareem A. and {Ukhorskiy}, Aleksandr Y. and {Merkin}, Viacheslav G. and {Fennell}, Joseph F. and {Claudepierre}, Seth G.},
        title = "{Modeling the Depletion and Recovery of the Outer Radiation Belt During a Geomagnetic Storm: Combined MHD and Test Particle Simulations}",
      journal = {Journal of Geophysical Research (Space Physics)},
         year = 2018,
        month = jul,
       volume = {123},
       number = {7},
        pages = {5590-5609},
          doi = {10.1029/2018JA025506},
       adsurl = {https://ui.adsabs.harvard.edu/abs/2018JGRA..123.5590S}
}

@ARTICLE{stawarz24SSR,
       author = {{Stawarz}, J.~E. and {Mu{\~n}oz}, P.~A. and {Bessho}, N. and {Bandyopadhyay}, R. and {Nakamura}, T.~K.~M. and {Eriksson}, S. and {Graham}, D.~B. and {B{\"u}chner}, J. and {Chasapis}, A. and {Drake}, J.~F. and {Shay}, M.~A. and {Ergun}, R.~E. and {Hasegawa}, H. and {Khotyaintsev}, Yu. V. and {Swisdak}, M. and {Wilder}, F.~D.},
        title = "{The Interplay Between Collisionless Magnetic Reconnection and Turbulence}",
      journal = {Space Science Reviews},
         year = 2024,
        month = dec,
       volume = {220},
       number = {8},
          eid = {90},
        pages = {90},
          doi = {10.1007/s11214-024-01124-8},
archivePrefix = {arXiv},
       eprint = {2407.20787},
 primaryClass = {physics.space-ph},
       adsurl = {https://ui.adsabs.harvard.edu/abs/2024SSRv..220...90S}
}

@ARTICLE{turner21a,
       author = {{Turner}, Drew L. and {Cohen}, Ian J. and {Michael}, Adam and {Sorathia}, Kareem and {Merkin}, Slava and {Mauk}, Barry H. and {Ukhorskiy}, Sasha and {Murphy}, Kyle R. and {Gabrielse}, Christine and {Boyd}, Alexander J. and {Fennell}, Joseph F. and {Blake}, J. Bernard and {Claudepierre}, Seth G. and {Drozdov}, Alexander Y. and {Jaynes}, Allison N. and {Ripoll}, Jean-Fran{\c{c}}ois and {Reeves}, Geoffrey D.},
        title = "{Can Earth's Magnetotail Plasma Sheet Produce a Source of Relativistic Electrons for the Radiation Belts?}",
      journal = {Geophys.~Res.~Lett.},
         year = 2021,
        month = nov,
       volume = {48},
       number = {21},
          eid = {e95495},
        pages = {e95495},
          doi = {10.1029/2021GL095495},
       adsurl = {https://ui.adsabs.harvard.edu/abs/2021GeoRL..4895495T}
}

@ARTICLE{turner21b,
       author = {{Turner}, Drew L. and {Cohen}, Ian J. and {Bingham}, Samuel T. and {Stephens}, Grant K. and {Sitnov}, Mikhail I. and {Mauk}, Barry H. and {Denton}, Richard E. and {Leonard}, Trevor W. and {Fennell}, Joseph F. and {Blake}, J. Bernard and {Torbert}, Roy B. and {Burch}, James L.},
        title = "{Characteristics of Energetic Electrons Near Active Magnetotail Reconnection Sites: Tracers of a Complex Magnetic Topology and Evidence of Localized Acceleration}",
      journal = {Geophys.~Res.~Lett.},
         year = 2021,
        month = jan,
       volume = {48},
       number = {2},
          eid = {e90089},
        pages = {e90089},
          doi = {10.1029/2020GL090089},
       adsurl = {https://ui.adsabs.harvard.edu/abs/2021GeoRL..4890089T}
}

@ARTICLE{usanova22,
       author = {{Usanova}, M.~E. and {Ergun}, R.~E.},
        title = "{Electron Energization by High-Amplitude Turbulent Electric Fields: A Possible Source of the Outer Radiation Belt}",
      journal = {Journal of Geophysical Research (Space Physics)},
         year = 2022,
        month = jul,
       volume = {127},
       number = {7},
          eid = {e30336},
        pages = {e30336},
          doi = {10.1029/2022JA030336},
       adsurl = {https://ui.adsabs.harvard.edu/abs/2022JGRA..12730336U}
}

@ARTICLE{wilson12GRL,
       author = {{Wilson}, III, L.~B. and {Koval}, A. and {Szabo}, A. and {Breneman}, A. and {Cattell}, C.~A. and {Goetz}, K. and {Kellogg}, P.~J. and {Kersten}, K. and {Kasper}, J.~C. and {Maruca}, B.~A. and {Pulupa}, M.},
        title = "{Observations of electromagnetic whistler precursors at supercritical interplanetary shocks}",
      journal = {Geophys.~Res.~Lett},
         year = 2012,
        month = apr,
       volume = {39},
       number = {8},
          eid = {L08109},
        pages = {L08109},
          doi = {10.1029/2012GL051581},
       adsurl = {https://ui.adsabs.harvard.edu/abs/2012GeoRL..39.8109W}
}

@ARTICLE{xjzhang25,
       author = {{Zhang}, Xiao-Jia and {Artemyev}, Anton V. and {Li}, Xinlin and {Arnold}, Harry and {Angelopoulos}, Vassilis and {Turner}, Drew L. and {Shumko}, Mykhaylo and {Runov}, Andrei and {Mei}, Yang and {Xiang}, Zheng},
        title = "{Relativistic and Ultra-Relativistic Electron Bursts in Earth's Magnetotail Observed by Low-Altitude Satellites}",
      journal = {Geophys.~Res.~Lett.},
         year = 2025,
        month = jan,
       volume = {52},
       number = {2},
        pages = {2024GL113280},
          doi = {10.1029/2024GL113280},
archivePrefix = {arXiv},
       eprint = {2408.17299},
 primaryClass = {physics.space-ph},
       adsurl = {https://ui.adsabs.harvard.edu/abs/2025GeoRL..5213280Z}
}


\end{document}